\documentclass[twoside,fleqn]{article}
\usepackage{espcrc2,amssymb,hyperref}
\usepackage[dvips]{graphicx}

\title{Thermodynamics of free domain wall and overlap fermions}

\author{
  George~T.~Fleming
  \address{
    Physics Department,
    The Ohio State University,
    Columbus OH 43210-1168
  }
}

\begin{document}

\pagestyle{empty}

\begin{abstract}

Studies of non-interacting lattice fermions give an estimate of the size
of discretization errors and finite size effects for more interesting problems
like finite temperature QCD.  We present a calculation
of the thermodynamic equation of state for free domain wall
and overlap fermions.

\end{abstract}

\maketitle

%%%%%%%%%%%%%%%%%%%%%%%%%%%%%%%%%%%%%%%%%%%%%%%%%%%%%%%%%%%%%%%%%%%%%%%%%%%%%%%%
\section{Introduction}
\label{sec:introduction}
%%%%%%%%%%%%%%%%%%%%%%%%%%%%%%%%%%%%%%%%%%%%%%%%%%%%%%%%%%%%%%%%%%%%%%%%%%%%%%%%

The thermodynamics of fermionic systems are certainly affected
by lattice regularization in a finite volume.  In numerical calculations of
the QCD equation of state at currently accessible lattice spacings, the data
tend to overshoot the continuum ideal gas limit.  This general pattern
is consistent with long established results \cite{Engels:1982ab}
for non-interacting lattice fermions.  This suggests that for any new lattice
fermion formulation, one should first study the thermodynamics
of the non-interacting system to gain insight on the potential cutoff effects
to be expected when the interactions are turned on.  For a recent discussion
and an example of this approach, see the review talk of Karsch from LATTICE '99
\cite{Karsch:2000vy}.

%%%%%%%%%%%%%%%%%%%%%%%%%%%%%%%%%%%%%%%%%%%%%%%%%%%%%%%%%%%%%%%%%%%%%%%%%%%%%%%%
\section{Energy density of free fermions}
\label{sec:eps_SB_disc}
%%%%%%%%%%%%%%%%%%%%%%%%%%%%%%%%%%%%%%%%%%%%%%%%%%%%%%%%%%%%%%%%%%%%%%%%%%%%%%%%

The energy density of a relativistic gas of non-interacting free fermions is
given by the integral
\begin{equation}
\label{eq:eps_SB}
\varepsilon_{\rm SB}(m) = \frac{2}{\pi^2} \int_0^\infty
\frac{\sqrt{p^2+m^2}}{1+\exp(\sqrt{p^2+m^2}/T)} p^2 {\rm d}p
\end{equation}
which is known analytically in the massless limit:
$\varepsilon_{\rm SB}(0) = ( 7 \pi^2 / 60 ) T^4$.  This is often called
the Stefan-Boltzmann (SB) limit from the $T^4$ dependence.
One important feature of this calculation is that a continuous set of momenta
are summed to compute the energy density.  However, similar calculations
for free fermions on the lattice are summed only over the finite set
of discrete lattice momenta. So, before we judge the relative merits of various
free lattice actions based on comparisons with the SB limit,
it is useful to first understand consequences of restricting free fermions
with a continuum dispersion relation to the allowed lattice momenta.

%%%%%%%%%%%%%%%%%%%%%%%%%%%%%%%%%%%%%%%%%%%%%%%%%%%%%%%%%%%%%%%%%%%%%%%%%%%%%%%%
\begin{figure}[h]
\includegraphics[width=0.46\textwidth]{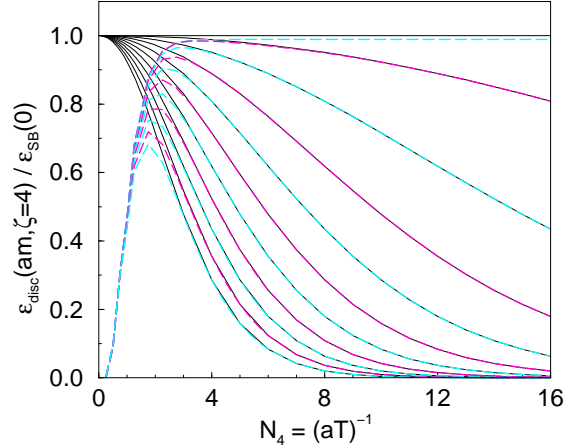}
\vspace{-1cm}
\caption{
  Discretized energy density of free fermions with $am=0,0.1,\cdots,1$.
}
\label{fig:one}
\end{figure}
%%%%%%%%%%%%%%%%%%%%%%%%%%%%%%%%%%%%%%%%%%%%%%%%%%%%%%%%%%%%%%%%%%%%%%%%%%%%%%%%

As this is a free field calculation, the scale is set explicitly by specifying
the dimensionful temperature $T$.  The infrared cutoff given by the spatial
extent of the lattice can be specified in terms of the dimensionless
{\it aspect ratio}: $\zeta\equiv VT^3$.  The ultraviolet cutoff given
by the lattice spacing can by specified in terms of the dimensionless
$N_4\equiv(aT)^{-1}$.  If we choose $a, V$ so that the product $\zeta N_4$
is an integer, then the discretized analogue of (\ref{eq:eps_SB}) is
\begin{equation}
\label{eq:eps_disc}
\frac{\varepsilon_{\rm disc}(m)}{T^4} = \frac{4}{\zeta^3} \sum_p
\frac{aN_4\sqrt{p^2+m^2}}{1+\exp(aN_4\sqrt{p^2+m^2})}
\end{equation}
where the sum is over $-\pi/a \le p < \pi/a$ with spacing
$\Delta p = 2 \pi / \zeta N_4 a$.

In figure \ref{fig:one}, we display the both the SB (\ref{eq:eps_SB}) and
discretized (\ref{eq:eps_disc}) energy densities at fixed aspect ratio
$\zeta=4$ for various $N_4$ and masses from $am=0$ to 1 spaced
in 0.1 intervals.  For high temperatures (small $N_4$), the mass
of the fermions becomes irrelevant and the SB (solid) lines converge
on $\varepsilon_{\rm SB}(0)$.  In this same limit, the number
of allowed lattice momenta becomes small so the discretized (dashed) lines
are cutoff and converge on zero.  For this aspect ratio, the SB and discretized
values differ by more than a percent for $N_4 \le 4$.

%%%%%%%%%%%%%%%%%%%%%%%%%%%%%%%%%%%%%%%%%%%%%%%%%%%%%%%%%%%%%%%%%%%%%%%%%%%%%%%%
\begin{figure}[t]
\includegraphics[width=0.46\textwidth]{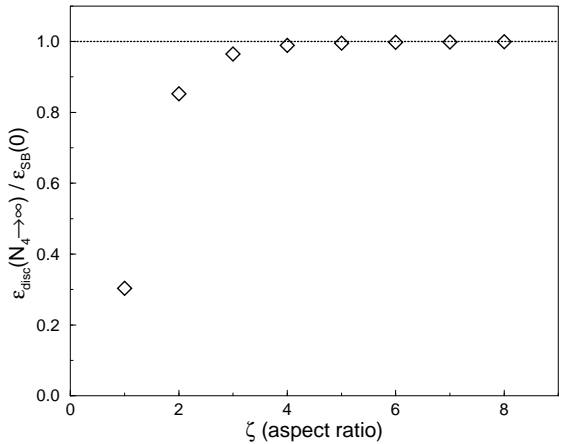}
\vspace{-1cm}
\caption{
  Discretized massless energy density {\it vs}.\ aspect ratio.
}
\label{fig:two}
\end{figure}
%%%%%%%%%%%%%%%%%%%%%%%%%%%%%%%%%%%%%%%%%%%%%%%%%%%%%%%%%%%%%%%%%%%%%%%%%%%%%%%%

In figure \ref{fig:one}, the massless discretized energy density falls short
of the SB limit, approaching 0.989 as $N_4\to\infty$.  For massless fermions,
even small momenta contribute at all temperatures, and these are cutoff
by the aspect ratio $\zeta=4$.  In figure \ref{fig:two}, we show the massless
discretized energy density {\it vs}.\ aspect ratio in the $N_4\to\infty$
limit.  So, from just the continuum free dispersion relation and a simple
discretization procedure, lattice simulations should not expect
to reproduce the continuum energy density to better than a few percent
on any lattice smaller than $16^3\times 4$.

%%%%%%%%%%%%%%%%%%%%%%%%%%%%%%%%%%%%%%%%%%%%%%%%%%%%%%%%%%%%%%%%%%%%%%%%%%%%%%%%
\section{Comparing free Wilson fermions}
\label{sec:Wilson_comparison}
%%%%%%%%%%%%%%%%%%%%%%%%%%%%%%%%%%%%%%%%%%%%%%%%%%%%%%%%%%%%%%%%%%%%%%%%%%%%%%%%

The energy density of free Wilson fermions has been known for a long time
\cite{Engels:1982ab} and its large overestimation of the SB limit at smaller
values of $N_4$ has often been used as an argument against the applicability
of Wilson thermodynamics at currently accessible lattice spacings.  Because
standard domain wall and overlap fermions use Wilson terms to give mass
to the doubler species, we might anticipate that similar effects may be seen
in the energy density of these fermion formulations.  So, we would like
to first understand free Wilson thermodynamics.

%%%%%%%%%%%%%%%%%%%%%%%%%%%%%%%%%%%%%%%%%%%%%%%%%%%%%%%%%%%%%%%%%%%%%%%%%%%%%%%%
\begin{figure}[t]
\includegraphics[width=0.46\textwidth]{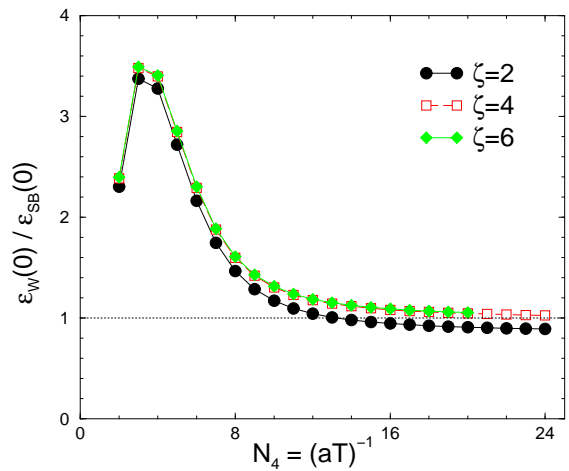}
\vspace{-1cm}
\caption{
  Energy density of free Wilson fermions.
}
\label{fig:three}
\end{figure}
%%%%%%%%%%%%%%%%%%%%%%%%%%%%%%%%%%%%%%%%%%%%%%%%%%%%%%%%%%%%%%%%%%%%%%%%%%%%%%%%

We show the energy density of free massless Wilson fermions {\it vs}.\ $N_4$
for three aspect ratios in figure \ref{fig:three}.  As in the discretized case
of section \ref{sec:eps_SB_disc}, approaching the $N_4\to\infty$ limit
at fixed small aspect ratio underestimates the SB result.  However,
the most obvious feature is the large overestimation of the SB result
at smaller $N_4$.  At the smallest $N_4$, at least, we now can understand
that the energy density is cutoff by the scarcity of lattice momenta.

%%%%%%%%%%%%%%%%%%%%%%%%%%%%%%%%%%%%%%%%%%%%%%%%%%%%%%%%%%%%%%%%%%%%%%%%%%%%%%%%
\begin{figure}[t]
\includegraphics[width=0.46\textwidth]{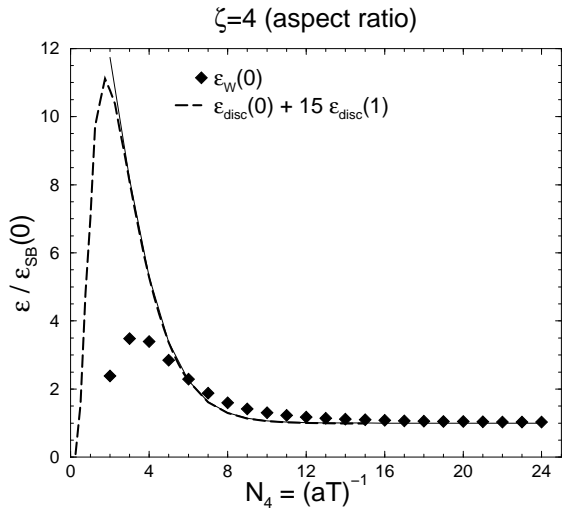}
\vspace{-1cm}
\caption{
  Discretized energy density with fifteen ``doublers.''
}
\label{fig:four}
\end{figure}
%%%%%%%%%%%%%%%%%%%%%%%%%%%%%%%%%%%%%%%%%%%%%%%%%%%%%%%%%%%%%%%%%%%%%%%%%%%%%%%%

We propose a simple model to reveal the source of the overestimate 
for Wilson fermions.  Without the Wilson term, the energy density
of free naive lattice fermions would be 16 times larger than the SB result.
The Wilson term gives 15 doublers a mass proportional to the cutoff.  But
from figure \ref{fig:one}, we expect these now massless doublers to make
a significant contribution to the overall Wilson energy density until
$N_4 \gtrsim 8$.  To emphasize the point, in figure \ref{fig:four}
we compare the free Wilson energy density with that of one massless
and fifteen massive flavors using the discretized energy density
(\ref{eq:eps_disc}).  While more effort could be made to adjust the masses
of the discretized doublers to make the two curves agree, from the common shape
we believe this is a good explanation of the Wilson excess.

%%%%%%%%%%%%%%%%%%%%%%%%%%%%%%%%%%%%%%%%%%%%%%%%%%%%%%%%%%%%%%%%%%%%%%%%%%%%%%%%
\section{Anisotropic domain wall action}
\label{sec:aniso_action}
%%%%%%%%%%%%%%%%%%%%%%%%%%%%%%%%%%%%%%%%%%%%%%%%%%%%%%%%%%%%%%%%%%%%%%%%%%%%%%%%

The energy density is formally given by
\begin{equation}
\label{eq:energy_density}
-\varepsilon \equiv \frac{1}{V} \left.
  \frac{\partial\ln Z}{\partial T^{-1}}
\right|_V .
\end{equation}
In order to vary the temperature continuously while holding the volume fixed
to compute the derivative, we must allow the lattice spacing in the temporal
direction, $a_4$, to differ from the spatial direction, $a$.  The ratio
of the spacings is the {\it anisotropy} $\xi\equiv a / a_4$.  Of course,
once the derivative has been computed we can take the isotropic limit,
$\xi\to 1$.

To derive the form of the anisotropic domain wall and overlap fermion matrix,
we choose the interpretation of the extra dimension in the domain wall
formalism as a flavor space.  We start from Klassen's anisotropic Wilson action
\cite{Klassen:1999fh} without the anisotropic clover terms
as the domain wall action is ${\cal O}(a)$ improved in the $L_s\to\infty$
limit.  Using the conventions of the Columbia group \cite{Chen:2000zu}
the anisotropic domain wall action is
\begin{equation}
\label{eq:dwf_dirac_matrix}
D_F(x,s;x^\prime,s^\prime) = \delta_{s,s^\prime} D^\parallel(x,x^\prime)
+D^\perp(s,s^\prime) \delta_{x,x^\prime}
\end{equation}
\begin{eqnarray}
\label{eq:dwf_dirac_matrix_parallel}
D^\parallel(x,x^\prime) = \sum_{\mu=1}^4 \xi^{z_\mu} \frac{\nu_\mu}{2}
\left[
  \left(1-\gamma_\mu\right) U_\mu(x) \delta_{x+\hat\mu,x^\prime}
\right. && \nonumber \\*
+ \left.
  \left(1+\gamma_\mu\right) U_\mu^\dagger(x^\prime) \delta_{x-\hat\mu,x^\prime}
\right] && \nonumber \\*
+ \left[ m_0 - r \left( 3 \xi^z \nu + \xi^{z_4} \nu_4 \right) \right] \quad &&
\end{eqnarray}
where the $\nu_\mu$ are the bare velocities of light
with $\nu_1=\nu_2=\nu_3\equiv\nu$, the anisotropy is $\xi\equiv a/a_4$,
and the exponents $z_\mu$ are set to $z_1=z_2=z_3\equiv z=0$ and $z_4=1$
although other choices are possible if they satisfy the constraint $z_4-z=1$
\cite{Trinchero:1983wc}.  The flavor mixing matrix $D^\perp(s,s^\prime)$ is
the same as written by the Columbia group \cite{Chen:2000zu}
without any extra coefficients since the normalization of the fermion fields
was shifted so that no coefficients appear with the $m_0$ term in the action.
The domain wall action includes a Pauli-Villars subtraction at $m_f=1$
to cancel the bulk infinity from the $L_s$ heavy flavors
at the mass scale $m_0$.

Following Neuberger \cite{Neuberger:1998bg}, the anisotropic form
of the overlap action can be derived as the $L_s\to\infty$ limit
of the anisotropic domain wall action.  However, the form of the action is
easy to guess:  just replace the isotropic Wilson matrix
with the anisotropic one in the overlap Hamiltonian.

%%%%%%%%%%%%%%%%%%%%%%%%%%%%%%%%%%%%%%%%%%%%%%%%%%%%%%%%%%%%%%%%%%%%%%%%%%%%%%%%
\section{Domain wall/overlap energy density}
\label{sec:eps_dwf}
%%%%%%%%%%%%%%%%%%%%%%%%%%%%%%%%%%%%%%%%%%%%%%%%%%%%%%%%%%%%%%%%%%%%%%%%%%%%%%%%

From (\ref{eq:energy_density}), the energy density is computed explicitly
using
\begin{equation}
\label{eq:energy_density_lat}
\frac{\varepsilon_F}{T^4} = \left(\frac{1}{\xi\zeta}\right)^3
{\rm Tr~} \left[
D_F^{-1} \xi \frac{\partial}{\partial\xi} D_F
\right] .
\end{equation}
Our approach was to compute the derivative $\xi \partial / \partial\xi D_F$
analytically and then diagonalize (\ref{eq:energy_density_lat}) into blocks
indexed by four-momentum $p_\mu$ using Fourier transform.  Each block matrix
$\widetilde{D}_F(p)$ was inverted numerically and then the traces were summed
over all four-momenta.  In general, (\ref{eq:energy_density_lat}) is computed
twice, once on a $(\zeta N_4)^3 \times N_4$ lattice and again
on a $(\zeta N_4)^4$ lattice for the $T=0$ vacuum subtraction.
Anti-periodic boundary conditions were imposed in the temporal direction
in both cases.  For domain wall fermions, the calculation is repeated with
the Pauli-Villars mass and the difference is used, schematically
\(
\varepsilon_{\rm dwf} = [ \varepsilon_F(T) - \varepsilon_F(0) ]
- [ \varepsilon_{PV}(T) - \varepsilon_{PV}(0) ] .
\)
%

%%%%%%%%%%%%%%%%%%%%%%%%%%%%%%%%%%%%%%%%%%%%%%%%%%%%%%%%%%%%%%%%%%%%%%%%%%%%%%%%
\begin{figure}[t]
\includegraphics[width=0.46\textwidth]{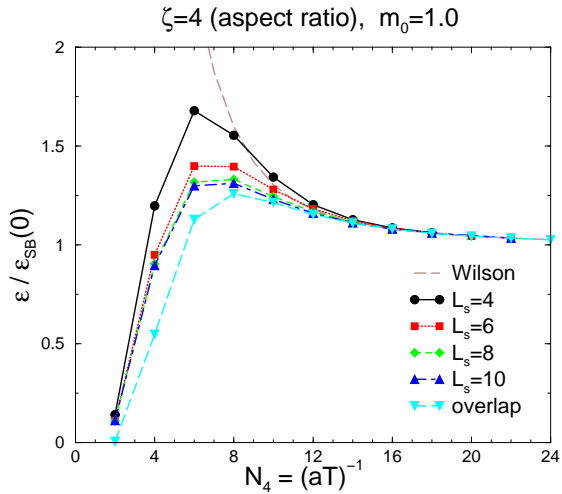}
\vspace{-1cm}
\caption{
  Domain wall and overlap energy density at $m_0=1$.
}
\label{fig:five}
\end{figure}
%%%%%%%%%%%%%%%%%%%%%%%%%%%%%%%%%%%%%%%%%%%%%%%%%%%%%%%%%%%%%%%%%%%%%%%%%%%%%%%%

The energy density for domain wall fermions at various $L_s$ and overlap
fermions are shown in figure \ref{fig:five}.  The choice $m_0=1$ is
the optimal free field value in the sense that the surface mode
with $p_\mu=0$ is completely localized to the domain wall.  However, surface
modes with other momenta will have finite exponential decay rates
so the $L_s$ dependence is due to these modes.  The energy density at $L_s=10$
is nearly converged on the $L_s\to\infty$ limit.  Any small differences
between the domain wall and overlap at smaller $N_4$ may be due to
the different treatment of the spacing of sites in the extra dimension.
This is currently under study.

%%%%%%%%%%%%%%%%%%%%%%%%%%%%%%%%%%%%%%%%%%%%%%%%%%%%%%%%%%%%%%%%%%%%%%%%%%%%%%%%
\begin{figure}[t]
\includegraphics[width=0.46\textwidth]{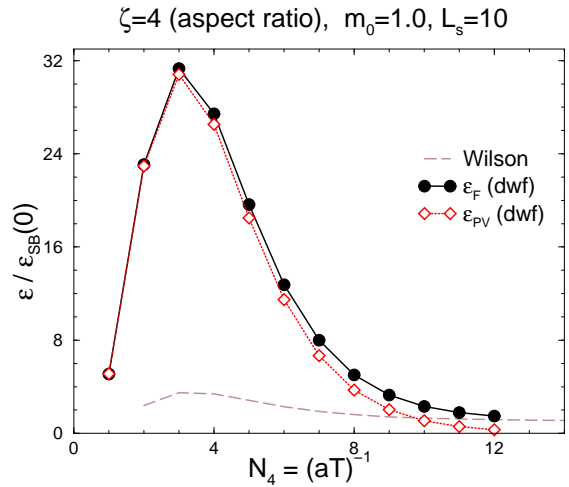}
\vspace{-1cm}
\caption{
  Domain wall fermion and Pauli-Villars energy densities.
}
\label{fig:six}
\end{figure}
%%%%%%%%%%%%%%%%%%%%%%%%%%%%%%%%%%%%%%%%%%%%%%%%%%%%%%%%%%%%%%%%%%%%%%%%%%%%%%%%

The Wilson energy density is also shown in figure \ref{fig:five}.
Based on arguments of section \ref{sec:Wilson_comparison}, the reduced excess
energy density for domain wall and overlap fermions indicates the participation
of the heavy doublers in the thermodynamics is suppressed
due to Pauli-Villars term.  In figure \ref{fig:six} we show the energy density
of domain wall fermions before Pauli-Villars subtraction and separately
the energy density of the Pauli-Villars regulators.  Before subtraction,
the excess energy density due to heavy modes is essentially $L_s$ times larger
than the Wilson case, again shown as a dashed line.  That the domain wall
density of order one appear as the difference of two densities of order $L_s$
suggests tinkering with the Pauli-Villars subtraction may reduce
heavy contributions further.  However, it is unclear how worthwhile a lengthy
study of possible Pauli-Villars terms would be when the benefits of such a
finely tuned action may vanish in the interacting case.

%%%%%%%%%%%%%%%%%%%%%%%%%%%%%%%%%%%%%%%%%%%%%%%%%%%%%%%%%%%%%%%%%%%%%%%%%%%%%%%%
\section{``Doubly'' regularized fermions}
\label{sec:doubly_reg_frm}
%%%%%%%%%%%%%%%%%%%%%%%%%%%%%%%%%%%%%%%%%%%%%%%%%%%%%%%%%%%%%%%%%%%%%%%%%%%%%%%%

As pointed out by Neuberger \cite{Neuberger:1998bg}, lattice fermion actions
with Pauli-Villars subtractions are ``doubly'' regularized, once
by the lattice spacing and again by the Pauli-Villars pseudofermions.
In section \ref{sec:eps_dwf}, we claim the Pauli-Villars suppression
of heavy contributions to the domain wall energy density works so well
the end result is an improvement over Wilson fermions quite apart
from any issues of chiral symmetry.  However, there is no reason we can't
attempt to use a Pauli-Villars regulator to achieve the same result
for Wilson fermions.

%%%%%%%%%%%%%%%%%%%%%%%%%%%%%%%%%%%%%%%%%%%%%%%%%%%%%%%%%%%%%%%%%%%%%%%%%%%%%%%%
\begin{figure}[t]
\includegraphics[width=0.46\textwidth]{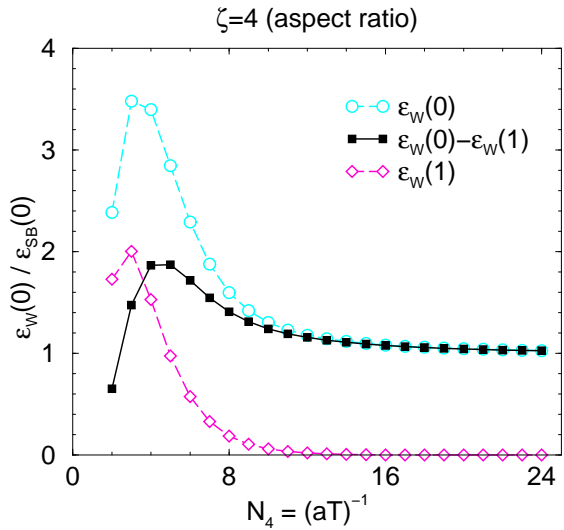}
\vspace{-1cm}
\caption{
  Energy density of ``doubly'' regularized Wilson fermions.
}
\label{fig:seven}
\end{figure}
%%%%%%%%%%%%%%%%%%%%%%%%%%%%%%%%%%%%%%%%%%%%%%%%%%%%%%%%%%%%%%%%%%%%%%%%%%%%%%%%

In figure \ref{fig:seven} we show the Wilson energy density
before Pauli-Villars subtraction, as has already appeared in the last several
figures as open circles, the energy density of the proposed Pauli-Villars
regulators with a mass of $m_0=1$ as open diamonds and the doubly regularized
Wilson energy density appears as filled squares.  After the subtraction,
the contribution of heavy modes to energy density is substantially reduced.
Fine tuning of the Pauli-Villars mass to make the energy density
agree with the Stefan-Boltzmann limit at a given $N_4$ is certainly possible,
but it is again unclear how to do this in the interacting case.

\begin{figure}[t]
\includegraphics[width=0.46\textwidth]{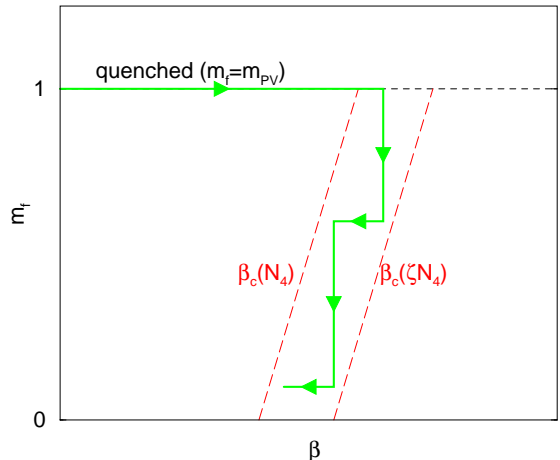}
\vspace{-1cm}
\caption{
  Integration contour for deconfined equation of state
}
\label{fig:eight}
\end{figure}

Another potential advantage of doubly regularized fermion formulations is
that the quenched theory is recovered when the fermion mass is set
to the Pauli-Villars mass.  An example of its usefulness is a proposed
calculation of the dynamical QCD equation of state in the deconfined phase
using the integral method.  As the integration contour originates
in the strong coupling regime and crosses through the transition region,
an inordinate amount of work must be done before anything is known
of the deconfined region.  As the quenched theory appears as a different
contour of fixed finite mass, one could use the integration contour shown
in figure \ref{fig:eight}.  As the quenched equation of state is already
well known for several $N_4$ \cite{Karsch:2000vy}, all dynamical simulations
are performed only at weaker couplings where fermion algorithms are usually
better conditioned.

%%%%%%%%%%%%%%%%%%%%%%%%%%%%%%%%%%%%%%%%%%%%%%%%%%%%%%%%%%%%%%%%%%%%%%%%%%%%%%%%
%%%%%%%%%%%%%%%%%%%%%%%%%%%%%%%%%%%%%%%%%%%%%%%%%%%%%%%%%%%%%%%%%%%%%%%%%%%%%%%%

One subtlety of integrating "off the quenched contour" is the importance
of computing the $T=0$ vacuum subtraction at as low a temperature as possible.
As before, finite temperature observables are computed
on $(\zeta N_4)^3 \times N_4$ lattices and $T=0$ subtractions are computed
on $(\zeta N_4)^4$ lattices.  On the quenched contour $(aT)^{-1}=\zeta N_4$
lattices may be confined, but while integrating down at fixed coupling,
one may enter the $\zeta N_4$ transition region.  To avoid this, a stair
step contour could be used.

We would like to thank N.~Christ, R.~Mawhinney, T.~Klassen and P.~Vranas
for useful discussions in the early stages of this work.

%%%%%%%%%%%%%%%%%%%%%%%%%%%%%%%%%%%%%%%%%%%%%%%%%%%%%%%%%%%%%%%%%%%%%%%%%%%%%%%%

%%%%%%%%%%%%%%%%%%%%%%%%%%%%%%%%%%%%%%%%%%%%%%%%%%%%%%%%%%%%%%%%%%%%%%%%%%%%%%%%

\end{document}